%% file: main.tex
\documentclass[conference]{IEEEtran}
\IEEEoverridecommandlockouts
\usepackage{subcaption}
\usepackage{fancyhdr,setspace}
\usepackage{comment}
\usepackage{caption}
\usepackage{todonotes}
\usepackage{xspace}
\usepackage{bm}

\usepackage[compact]{titlesec} 
\usepackage{cite}
\usepackage{blindtext, graphicx, tabularx, multicol, lipsum, subfig,eqnarray,amsmath,chemformula,multirow,threeparttable}
\usepackage{enumitem}
\usepackage[utf8]{inputenc}
\usepackage{amssymb,amsmath}

\usepackage{float}

\usepackage[boxed]{algorithm2e}
\usepackage[noend]{algpseudocode}
\usepackage[compact]{titlesec}
\algnewcommand{\IfThenElse}[3]{
  \State \algorithmicif\ #1\ \algorithmicthen\ #2\ \algorithmicelse\ #3}
\makeatletter
\newcommand{\removelatexerror}{\let\@latex@error\@gobble}
\makeatother
\makeatletter
\renewcommand\footnoterule{%
  \kern-3\p@
  \hrule\@width\columnwidth
  \kern2.6\p@}
  \makeatother
\usepackage{letltxmacro}

\usepackage{amsthm}

\usepackage{amsmath}

\SetKwInput{KwInput}{Input}                
\SetKwInput{KwOutput}{Output}  

\usepackage{tikz}
\usepackage{capt-of}
\usepackage{amssymb} \usepackage{booktabs} 
\usepackage{pifont}
\newcommand{\xmark}{\ding{55}}%

\usepackage{diagbox}
\usepackage{xcolor,etoolbox}
\usepackage{dblfloatfix}

\let\mybibitem\bibitem
\renewcommand{\bibitem}[1]{%
\ifstrequal{#1}{edgeTPU}{\color{black}\mybibitem{#1}}
{\ifstrequal{#1}{xyz}{\color{blue}\mybibitem{#1}}
{\color{black}\mybibitem{#1}}}%
}

\ifCLASSINFOpdf
\else
\fi
\usepackage{xcolor}
\usepackage{comment}
\usepackage{glossaries}
\input{glossar}

\usepackage{fancyhdr}
\renewcommand{\headrulewidth}{0pt}
\begin{document}



\title{SPICED: \textbf{S}yntactical Bug and Trojan \textbf{P}attern \textbf{I}dentification in A/MS \textbf{C}ircuits using LLM-\textbf{E}nhanced \textbf{D}etection}

\author{
\IEEEauthorblockN{Jayeeta Chaudhuri$\dagger$, Dhruv Thapar$\dagger$, Arjun Chaudhuri$\ddagger$, Farshad Firouzi$\dagger$, and Krishnendu Chakrabarty$\dagger$}
\IEEEauthorblockA{$\dagger$School of Electrical, Computer, and Energy Engineering, Arizona State University, AZ, USA\\
$\ddagger$NVIDIA Corporation, CA, USA}
}

\maketitle

\begin{abstract}

Analog and mixed-signal (A/MS) integrated circuits (ICs) are crucial in modern electronics, playing key roles in signal processing, amplification, sensing, and power management. Many IC companies outsource manufacturing to third-party foundries, creating security risks such as stealthy analog Trojans. Traditional detection methods, including embedding circuit watermarks or conducting hardware-based monitoring, often impose significant area and power overheads, and may not effectively identify all types of Trojans. To address these shortcomings, we propose SPICED, a Large Language Model (LLM)-based framework that operates within the software domain, eliminating the need for hardware modifications for Trojan detection and localization. This is the first work using LLM-aided techniques for detecting and localizing syntactical bugs and analog Trojans in circuit netlists, requiring no explicit training and incurring zero area overhead. Our framework employs chain-of-thought reasoning and few-shot examples to teach anomaly detection rules to LLMs. With the proposed method, we achieve an average Trojan coverage of 93.32\% and an average true positive rate of 93.4\% in identifying Trojan-impacted nodes for the evaluated analog benchmark circuits. These experimental results validate the effectiveness of LLMs in detecting and locating both syntactical bugs and Trojans within analog netlists. 

\end{abstract}

\thispagestyle{fancy}
\fancyhead{}
\renewcommand{\headrulewidth}{0pt}
\fancyhf{}
\fancyfoot[C]{\thepage}


%
\IEEEpeerreviewmaketitle

\input{Introduction}


\input{Background}
\input{Bug_detection}


\input{Proposed_framework}

\input{Results}

\input{Conclusion}


\bibliographystyle{myIEEEtran}
\bibliography{references}

\end{document}

%% file: glossar.tex
\newacronym{eda}{EDA}{Electronic Design Automation}

\newacronym{hdl}{HDL}{Hardware Description Language}

\newacronym{llm}{LLM}{Large Language Model}

\newacronym{rtl}{RTL}{Register-Transfer Level}

\newacronym{gds}{GDSII}{Graphic Data System Version II}

%% file: Introduction.tex
\section{Introduction}

Analog and mixed-signal (A/MS) integrated circuits (ICs) play a critical role in signal processing, amplifiers, sensors, and power management systems. Many IC companies such as Intel, AMD, Qualcomm, and Texas Instruments opt to outsource the manufacturing and fabrication of their analog designs to third-party foundries to avoid expensive capital expenditures as well as substantial costs associated with investing in manufacturing infrastructure. The globalization of the semiconductor industry and the outsourcing of analog ICs to third-party vendors have introduced significant security threats. These threats notably compromise the integrity of ICs, making them vulnerable to analog Trojans \cite{alam2018challenges} \cite{6860363} \cite{9794141}. A major risk arises from the possibility of embedding stealthy Trojans that evade detection under normal operating conditions \cite{7546493} \cite{10.1145/3489517.3530666}. 


Trojans occupy minimal area footprints, enabling their easy integration into larger, complex A/MS designs at the netlist level, which includes multiple paths and transistor components. These stealthy components are activated only during specific operating bias voltages and remain dormant otherwise. Prior work on analog Trojan detection utilizes current-sensing amplifiers to identify Trojan activation \cite{9937796}. A recent study introduces a sensitivity analysis-based framework leveraging analog neural twins to detect stealthy analog Trojans \cite{9983931}. This approach identifies the critical paths most vulnerable to Trojan insertion \cite{10490135}. Next, circuit watermarks are embedded to monitor deviations in these paths, triggering an alert when a Trojan is activated. While these methods are effective in detecting Trojans, they do not address the challenge of locating Trojan-impacted nodes within the design netlist, a task that becomes increasingly difficult as circuit complexity grows.


The recent advancements in Large Language Models (LLMs) have showcased their significant capabilities across various tasks including code generation and optimization. Such advancements in the EDA domain naturally position LLMs as highly potent for novel applications in A/MS design. Traditional methods for Trojan detection, including embedding circuit watermarks or conducting hardware-based monitoring, frequently impose significant area and power overheads, especially in large analog designs. To address this issue, we propose SPICED, an LLM-based framework that operates within the software domain, thereby eliminating the need for any hardware modifications to the analog design for Trojan detection and localization. SPICED excels at intelligent parsing and analysis of large volumes of structured data such as SPICE netlists. In addition to Trojan localization, the proposed LLM-based framework provides comprehensive analysis and detailed diagnosis of the detected anomalies. By leveraging a deep understanding of the HSPICE language, simulation logs, and the topological structure of netlists, SPICED can not only distinguish between Trojan-free and Trojan-inserted netlists but also precisely identify the specific Trojan components and the nodes affected by the Trojan. The key contributions of this paper are as follows:


\begin{itemize}[leftmargin=*,topsep=0pt]

    \item \textbf{Introduction of SPICED}: Presenting SPICED, the first LLM-based framework for Trojan detection in A/MS design that requires no hardware modifications.


    \item \textbf{Syntactical Bug Mitigation:} Leveraging in-context learning and Chain-of-Thought (CoT) prompting to detect and mitigate syntactical bugs in SPICE netlists.

    \item \textbf{Precise Trojan Detection and Localization: } Developing supervised learning rules for the LLM to identify Trojan circuits and the Trojan-affected nodes in an analog design.

    
\end{itemize}
  
The remainder of the paper is organized as follows. Section II provides a comprehensive overview of recent work on LLM-aided text generation and how its capability can be leveraged for bug detection tasks in the analog domain. We provide a detailed analysis of syntactical bug detection and correction using LLM in Section III. Section IV presents the framework for Trojan detection and localization using LLM involving both CoT and few-shot prompting examples for generating supervised learning rules. Evaluation results for SPICED with comparisons among several LLM-aided techniques are presented in Section V. Finally, Section VI concludes the paper.

%% file: Background.tex
\section{Background and Motivation}




\begin{table}[t]
\centering
\fontsize{7.8}{7.8}\selectfont
\caption{{Comparison of prior works using LLM-aided techniques.}}

    \label{compare}
    \begin{tabular}{|c| c| c| c|c|c|} 
    \hline
{Method} & Domain&Training-&Bug &Bug&Trojan \\
&&Free?&Detection?&Fixing?&Detection\\
\hline
\cite{verigen}&Digital&\xmark& \xmark &\xmark&\xmark \\

\cite{amsnet}&Analog&\xmark&\xmark&\xmark&\xmark  \\

\cite{analogcoder}&Analog&\checkmark&\xmark&\xmark&\xmark  \\

\cite{chipchat}&Digital&\checkmark&\xmark&\xmark &\xmark\\

\cite{chipnemo}&Digital&\xmark&\xmark&\xmark&\xmark\\

\cite{thakur2023benchmarking}&Digital&\xmark&\xmark&\xmark&\xmark\\
\cite{bhandari2024sentaur}&Digital&\checkmark&\xmark&\xmark&\xmark \\
\cite{tsai2023rtlfixer}&Digital&\checkmark&\checkmark&\checkmark&\xmark \\
\textbf{SPICED }&Analog&\checkmark&\checkmark&\checkmark&\checkmark \\
\hline
\end{tabular}
\vspace{-0.2cm}
\label{comp}
\end{table}

\subsection{Design Automation using LLMs}

LLMs have been integrated into multiple stages of EDA, enhancing processes from code generation and placement and routing to security measures, thereby streamlining and improving the overall design workflow. Recent work used LLM for Verilog code generation by fine-tuning existing LLM with Verilog datasets \cite{verigen}. Fine-tuned open-source CodeGen LLM outperformed state-of-the-art commercial LLM in generating functionally correct designs \cite{thakur2023benchmarking}. \cite{chipchat} used LLM-based iterative flow to design an 8-bit accumulator-based microprocessor architecture. \cite{analogcoder} proposed LLM for generating analog circuits with a feedback-enhanced flow to enable self-correcting design of analog circuits. The feedback allowed generation of circuits without any LLM training involved. Domain-Adaptive Pre-Training (DAPT) followed by Supervised Fine-Tuning (SFT) of foundation LLM models enabled an assistant chatbot for chip design \cite{chipnemo}. \cite{chateda,llm4eda} proposed LLM-based script generation to facilitate the EDA design flow. In the context of security, LLM is shown to be effective in structural generation of digital Trojans \cite{bhandari2024sentaur} as well as fixing syntactical bugs in Verilog codes \cite{tsai2023rtlfixer}. These works are focused mainly on code generation tasks and bug detection in the digital domain, as shown in Table \ref{comp}. 
\vspace{-0.1cm}

\subsection{Nature of Analog Trojans}
\vspace{-0.1cm}
An analog Trojan consists of two primary components: (1) Trigger circuit, which is conditionally gated with an AND or OR gate to activate the Trojan only upon specific toggling instructions; (2) Detector circuit, which detects the charge buildup of the capacitive component of the Trojan circuit, and when the capacitor voltage reaches a threshold, activates the payload. Recent work shows the impact of an analog Trojan, namely A2 \cite{7546493} that can be stealthily inserted during the design and fabrication phases of an analog design. Due to its small footprint, it can be maliciously inserted in unused parts during design phase. The trigger for the A2 Trojan is software-controlled i.e., the trigger is activated when a rare instruction is executed. As an extension to the A2 Trojan, authors in \cite{10.1145/3489517.3530666} shows the implementation of the DELTA Trojan, which uses a glitch generator for the trigger circuit, and can be inserted in any net of a circuit irrespective of it being rarely activated. 

In \cite{10490135}, it has been demonstrated that A2 Trojans remain mostly dormant due to their insertion in the less sensitive paths of a circuit. Upon their activation, the primary output voltage behavior is impacted, leading to significant performance degradation. Therefore, it is necessary to detect and localize these malicious circuits within the analog design netlist before the netlist is sent to the fabrication stage.

\vspace{-0.1cm}

\subsection{Prior Work on Analog Trojan Detection}
\vspace{-0.1cm}
In \cite{9937796}, a current sensing-based circuit is inserted in a digital design to detect analog Trojans such as A2 at run-time. A recent work \cite{10490135} demonstrates the impact of analog Trojans (A2, DELTA, and large-delay Trojans) on A/MS designs as well. In \cite{10490135}, analog neural twins are employed to identify critical paths in an analog circuit netlist through sensitivity analysis. After identifying critical paths, circuit watermarks are inserted to make these paths observable at the circuit output. Sensitizing the least sensitive paths of a circuit makes the detection of stealthy Trojans easier. This ensures that even if the effects of the Trojans are not captured at the primary output of the circuit, they can still be detected through the altered behavior of the sensitized paths.

Although \cite{10490135} effectively identifies all the Trojan hotspots, there are two significant limitations: (1) the area overhead associated with added watermarks increases with the complexity of the analog design, and (2) the specific Trojan-inserted nodes are not identified, i.e., localization of the detected Trojans is not performed; this limitation makes it difficult to pinpoint the exact nodes affected by the Trojans, making targeted mitigation infeasible. 

These limitations motivate the exploration of specialized techniques that leverage the contextual ability of LLMs. LLMs are vastly known for interpreting and generating texts, and are capable of interpreting contexts across various programming languages such as C, C++, and Python. A significant portion of the training data for LLMs is sourced from Github repositories. LLMs such as Llama-2 and Llama-3 are primarily trained on Python datasets \cite{roziere2023code}. In the realm of digital design, the availability of Verilog code in training datasets is relatively limited compared to other programming languages \cite{analogcoder}. The scarcity is even more for analog design data, particularly for the SPICE language, which has less code available in open-source repositories. This poses a challenge for LLMs like Llama 2 and GPT-3.5 to understand SPICE syntax comprehensively. Despite these challenges, GPT-3.5 shows the capability of understanding basic SPICE syntax and type of circuit configuration being implemented in a SPICE netlist. Leveraging LLMs for design analysis offers several advantages.

\begin{itemize}[leftmargin=*,topsep=0pt]
\item \textbf{Textual analysis capability}: Aided by the right prompts, LLMs can effectively parse simulation log files without any manual intervention, thereby offering huge productivity boost in terms of reduced engineering hours and log processing runtime. LLMs understand the tabular structure of logged voltage and current values in the logs, making it easier for them to extract the numerical voltage and current data for each circuit node.

\item \textbf{Netlist Identification and Bug Correction}: Pre-trained LLMs can understand the basic syntax of SPICE netlists, enabling the models to analyze the netlist structure and flag syntax errors.

    \item \textbf{Anomaly detection}: LLMs can learn the node properties and inter-node relationships described by the netlist topology and simulation logs, and identify anomalous patterns among normal operational characteristics of the design via in-context learning.

    \item \textbf{Trojan Detection and Localization}: Equipped with the ability to parse logs, analyze netlists, and identify outlier patterns, LLMs can potentially identify Trojan-affected nodes whose voltage and current profiles manifest as anomalous. 

    \item \textbf{No Area and Power Overhead}:  Leveraging the computational abilities of LLMs obviates the need for any additional Trojan detection hardware in the circuit.
\end{itemize}

\vspace{-0.1cm}

%% file: Bug_detection.tex
\section{LLM for Syntactical Bug Detection}

\vspace{-0.1cm}

\begin{figure}[b]
\centering
\includegraphics[width=0.4\textwidth]{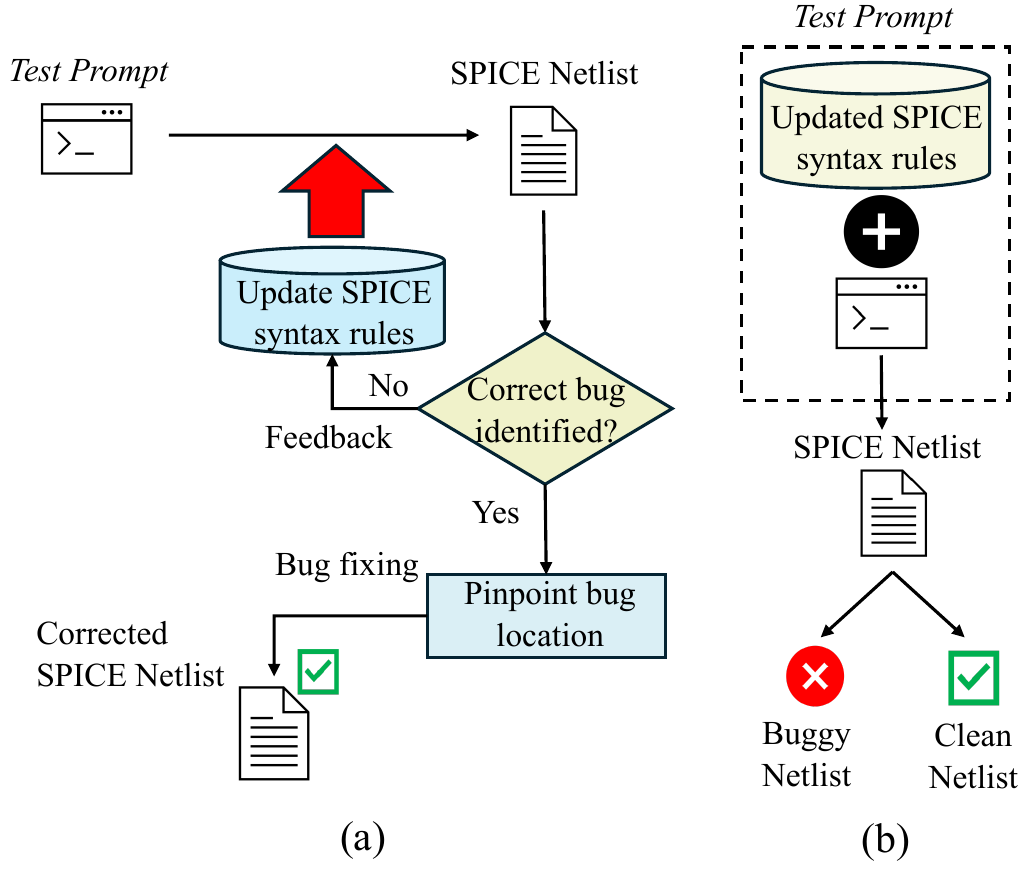}
  \caption{(a) Enhancing syntactical bug detection accuracy of LLM by instruction-following approach (b) Identifying and mitigating bugs in a SPICE netlist. }
  \label{bug_method}
\vspace{-0.5cm}
\end{figure}

\begin{figure*}
\centering
\includegraphics[width=\textwidth]{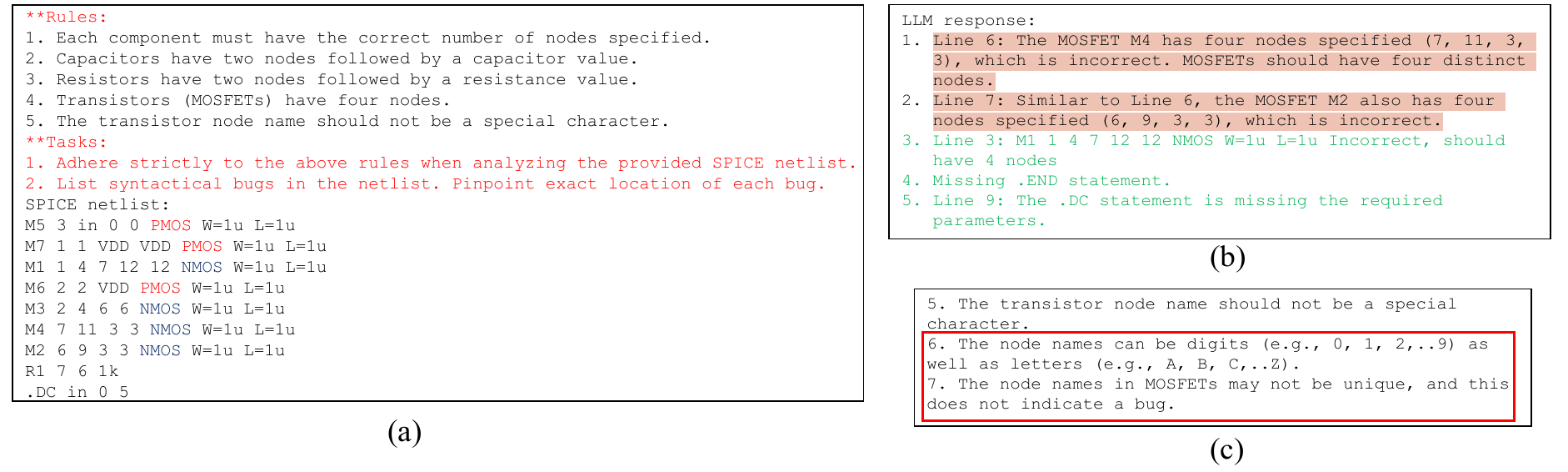}
  \caption{(a) An example prompt highlighting the SPICE syntax rules for bug detection and localization (b) LLM correctly detects the bugs injected in the netlist; however, it incorrectly flags Lines 6 and 7 as bugs (highlighted in red) (c) Explicitly updating the rules in the prompt to reduce false positives. }
  \label{instruct}
\vspace{-0.2cm}
\end{figure*}


 Fig. \ref{bug_method} illustrates the proposed flow of guiding the LLM to detect a wide range of syntactical bugs while reducing the likelihood of incorrectly flagging correct lines as syntax errors. Note that this method does not involve any LLM fine-tuning; instead, it focuses on improving the bug detection capability of LLM through refinement of SPICE syntax rules based on real-time feedback. The steps involved are as follows:

\begin{enumerate}[leftmargin=*,topsep=0pt]

    \item Begin with an initial set of SPICE syntax rules and use them to construct the LLM prompt to identify and locate syntactical bugs within a SPICE netlist (shown in Fig. \ref{instruct}(a)).
    \item When the LLM identifies the bug correctly, it specifies the type of the detected bug and its location in the netlist (shown in Fig. \ref{instruct}(b)). 
    \item A false positive occurs when the LLM incorrectly identifies a line as containing a bug. If the LLM incorrectly flags a line as buggy, this feedback is used to manually update and refine the syntax rules in the prompt (shown in Fig. \ref{instruct}(c)). 
    \item After updating the rules, the refined prompt is applied to the same netlist. The process is repeated until the number of false positives is minimized.
     
\end{enumerate}

We use the updated set of syntax rules for evaluation. Additionally, we prompt the LLM to generate a structured bug detection report. The report includes the following items: 
\begin{enumerate}
   [leftmargin=*,topsep=0pt]
  
    \item \emph{List of all syntactical bugs in the SPICE netlist}: The bugs may include connection error (missing connections or floating nodes in the netlist), insertion error (unintended or intended insertion of additional transistor components), and incorrect specifications of parameters.
    \item \emph{Location of the bugs in the SPICE netlist}: Lines in the SPICE netlist where the bug is located, including the component or connection names.
    \item \emph{Suggestions for correction}: Provides a list of recommended actions to correct the bugs and generate a revised netlist.
\end{enumerate}

\vspace{-0.2cm}

%% file: Proposed_framework.tex
\begin{figure*}
\centering
\includegraphics[width=\textwidth]{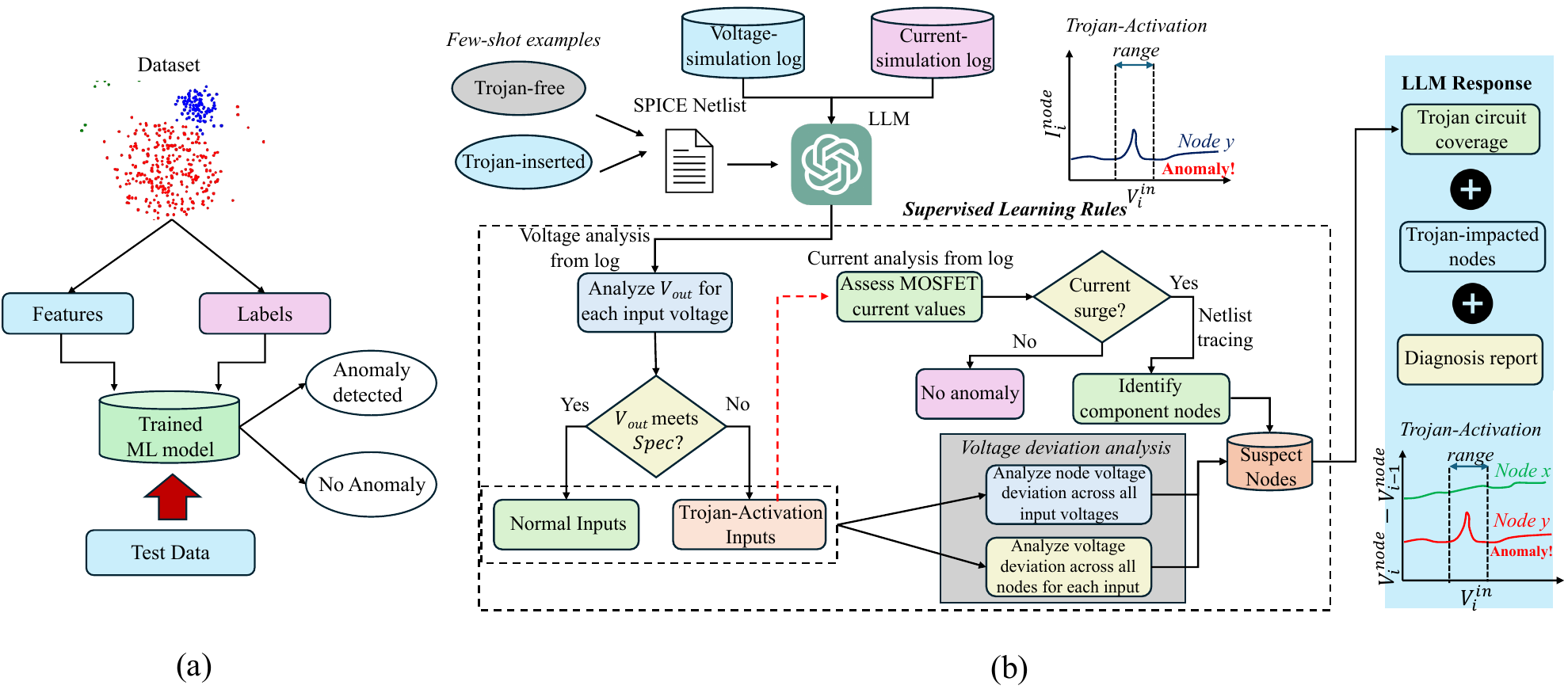}
  \caption{(a) Flow of supervised-learning approach using machine learning (ML) models (b) Using the supervised-learning analogy to locate Trojan-impacted nodes using LLM ($V_{out}$: primary output voltage, $Spec$: desired output voltage specifications, $V^{node}_i (I^{node}_i$): node voltage (current) corresponding to $i^{th}$ input voltage sample $V^{in}_i$). }
  \label{trojan_method}
\vspace{-0.4cm}
\end{figure*}

\section{Supervised Learning-Based Framework using LLM for Trojan Detection }

\vspace{-0.1cm}
While Section III focuses on identifying and correcting syntactical bugs in the SPICE netlist, this section addresses functional bug detection, particularly targeting analog Trojans. A Trojan-impacted node is defined as one of the circuit nodes where the Trojan is inserted or which is part of the trigger node.
 From prior work on analog Trojan detection \cite{10490135} \cite{9937796}, the following observations are noted regarding the current and voltage behavior of Trojan-impacted nodes.

\begin{enumerate}[leftmargin=*,topsep=0pt]

\item Deviation in primary output voltage: As observed in \cite{10490135}, when a Trojan is inserted into one of the sensitive paths of an analog design, its impact is captured at the primary output voltage. Specifically, the output voltage exceeds the desired specifications when the Trojan is activated.
    \item Anomalous deviation in circuit intermediate nodes: Nodes impacted by the Trojan, or the neighboring nodes exhibit significant deviations in voltage behavior when the Trojan is activated. This deviation can serve as a critical indicator of the presence of a Trojan.
    \item Anomalous surge in MOSFET current: According to \cite{9937796}, upon Trojan activation, some MOSFETs within the analog design draw a substantially higher current. This current behavior deviates significantly from the normal current behavior observed when the Trojan circuit is dormant.
\end{enumerate}
\begin{figure}
\centering
\includegraphics[width=0.42\textwidth]{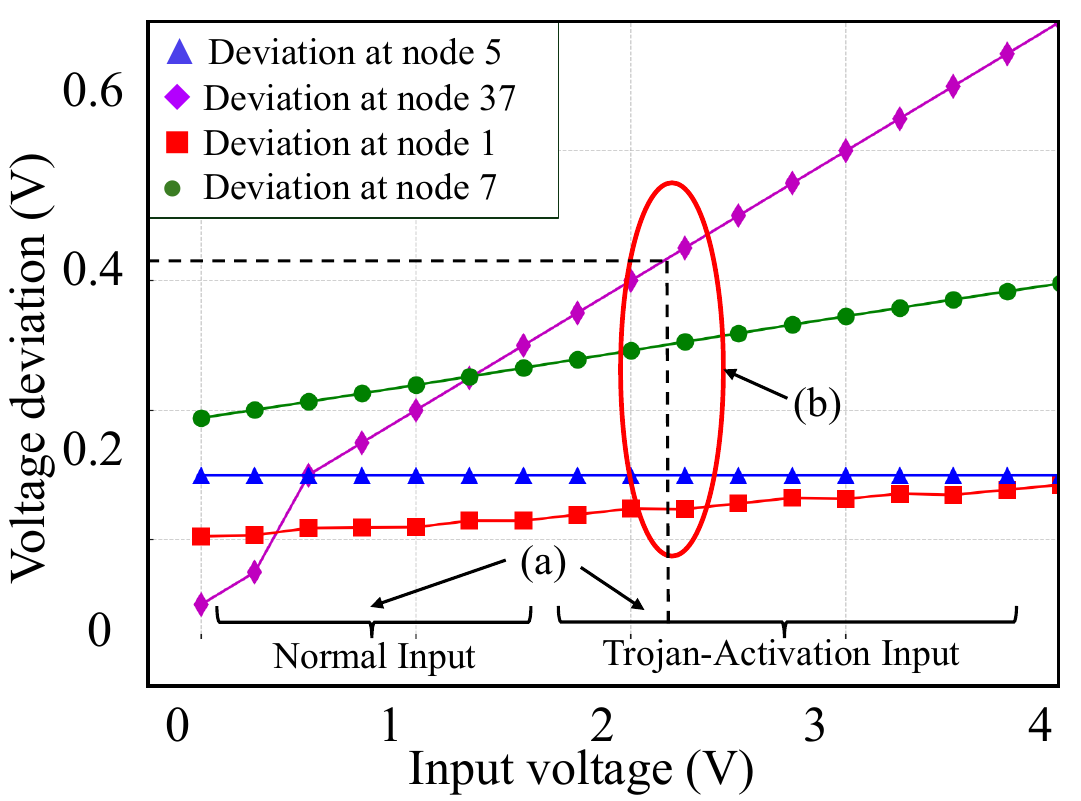}
  \caption{Voltage deviation analysis for A2 Trojan-inserted netlist of circuit `642' from AMSNet \cite{amsnet}. Voltage deviation of a node $x$ is given by $V_i^{x}-V_{i-1}^x$, where $i-1$ and $i$ are consecutive input voltage samples. (a) Analyze voltage deviation of each node across the `Normal Input' and `Trojan-Activation Input' ranges, (b) analyze voltage deviation across nodes in the `Trojan-Activation Input' range. Combining (a) and (b), we observe that node 37 is a Trojan-impacted node.}
  \label{trojan_plot}
\vspace{-0.5cm}
\end{figure}

For identifying analog Trojans and the Trojan-impacted nodes, we used these observations to craft supervised-learning rules. We then created prompts based on these rules to teach the LLM to detect anomalies in the current and voltage simulation logs. The simulation logs are obtained by running HSPICE simulation on analog netlists. The procedure of teaching the supervised learning approach (shown in Fig. \ref{trojan_method} (a)) to the LLM is shown in Fig. \ref{trojan_method}(b). We link each Trojan characteristic to a specific supervised learning rule as follows:

\textbf{1. Deviation in primary output voltage}

$\blacksquare$ \textbf{Supervised Learning Rule 1:} We craft prompts that highlight output voltages that exceed the desired circuit specifications. Based on these output voltages, we ask the LLM to identify the corresponding input voltages. These input voltages are labeled as `Trojan-Activation Inputs', while inputs where the Trojan is inactive are labeled as `Normal Inputs.'

\textbf{2. Anomalous deviation in circuit intermediate nodes}

$\blacksquare$ \textbf{Supervised Learning Rule 2:}  Fig. \ref{trojan_plot} shows the voltage deviations in intermediate nodes of a Trojan-inserted circuit across different input voltages and how their voltage deviations compare across nodes for anomaly detection. The LLM identifies the nodes that exhibit (a) a significant voltage deviation  between `Trojan-Activation Input' and `Normal Input' ranges, and (b) the highest deviation among all the intermediate nodes for each `Trojan-Activation Input'. Based on these observations, the LLM applies one of the following rules to determine the Trojan-impacted nodes.
\begin{itemize}[leftmargin=*,topsep=0pt]
    \item Rule \#1: Union of  nodes identified in steps (a) and (b).
    \item Rule \#2: Intersection of nodes obtained in steps (a) and (b).
\end{itemize}

\textbf{3. Anomalous surge of MOSFET current}

$\blacksquare$ 
 \textbf{Supervised Learning Rule 3:} We design prompts to examine current simulation logs and identify MOSFETs that generate unusual current spikes under `Trojan-Activation Inputs' compared to the baseline of `Normal Inputs'. The nodes corresponding to the MOSFETs with detected current anomalies are flagged as suspect nodes.

 Finally, combining the above supervised learning rules, we prompt the LLM to detect the Trojan circuit within an analog netlist and generate the final set of Trojan-impacted nodes. The supervised learning rules are defined at the start of the prompt. Next, we provide the LLM with few-shot examples, where it sees a limited number of example netlists, labeled as `Trojan' or `Trojan-Free'. Additionally, we provide explanations related to the supervised-learning rules and the simulation behavior of current and voltage from the log files to justify deviations that indicate potential Trojan behavior.

\vspace{-0.1cm}


%% file: Results.tex
\section{Experimental Results}
\vspace{-0.1cm}

\subsection{Experimental Setup}
\vspace{-0.1cm}
For both bug detection and Trojan detection experiments, we evaluate a wide variety of analog benchmark circuits that are selected from the OpenSource netlist dataset from AMSNet \cite{amsnet} and Github \cite{ads}. These circuits include differential amplifier, inverter, OPAMP, and bandpass filter, thus providing a comprehensive dataset for evaluating both syntactical bug and Trojan detection capabilities of the LLM. We have compiled a total of 18 syntactical bugs from \cite{10.5555/528264}. We choose varying complexity of bugs to test the effectiveness of LLM in identifying bug-inserted SPICE netlists. The bug complexity is classified as: easy, medium, and difficult. The bug benchmark used for LLM evaluation is shown in Table \ref{bugbench}. The benchmark includes 4 easy, 6 medium, and 8 difficult syntactical bugs that can be present in a SPICE netlist.

For the Trojan-detection experiments, we evaluate the well-known analog Trojan explored in recent literature - A2 \cite{7546493}. This type of Trojan occupies small footprint and hence, can be easily embedded in the netlist stage by an untrusted foundry. We embed A2 in the SPICE netlist of an analog benchmark circuit to generate a Trojan-inserted netlist. The Trojan is considered to be internally triggered by an intermediate node of the analog design, thus emulating a realistic attack scenario shown in \cite{5677557}. For our experiments, we use the GPT-3.5-turbo API. The experiments are carried out on an NVIDIA A100 GPU.
Table \ref{inputs} lists the information available to the LLM for performing bug and Trojan detection tasks.

\vspace{-0.1cm}

\subsection{Evaluation Metrics}

\vspace{-0.1cm}
We use the following metrics to evaluate the effectiveness of the LLM in syntactical bug and Trojan detection.

\begin{itemize}[leftmargin=*,topsep=0pt]
    
    \item Bug coverage (in \%): This metric represents the percentage of syntactical bugs detected by the LLM out of the total number of bugs embedded within a SPICE netlist. 
    \item Bug resolved (in \%): It represents the percentage of detected syntactical bugs that are correctly fixed by LLM.
    \item Trojan identified: It indicates whether the LLM has correctly detected at least one Trojan component in the netlist.
    \item Trojan coverage (in \%): This metric calculates the ratio of the number of malicious components (transistor, resistor, or capacitor) embedded in the SPICE netlist that are correctly identified by the LLM and the total number of Trojan-injected components in the netlist.
    
    \item Precision (in \%): It indicates the percentage of correctly predicted Trojan-impacted nodes out of all the predicted impacted nodes. It is denoted by $Precision = \frac{TP}{TP+FP}$, where $TP$ and $FP$ indicate the true positive and false positive counts, respectively.

    \item Recall (\%): It is indicated by the percentage of correctly predicted Trojan-impacted nodes out of the actual number of Trojan-impacted nodes. It is denoted by: $Recall = \frac{TP}{TP+FN}$, where $TP$ and $FN$ indicate the true positive and false negative counts, respectively.

\end{itemize}

\vspace{-0.1cm}
\begin{table}[t]
\centering
\caption{List of syntactical bugs with varying complexity for evaluating LLM performance.}
\fontsize{6.5}{6.5}\selectfont
\setlength\tabcolsep{3pt}
    
 \begin{tabular}{|c|c|c|}
     \cline{1-3}

Description of Bug& Example&Complexity\\
\hline
Missing node of transistor&M1 2 3 0 PMOS&E\\
\hline
Missing .END statement&Netlist without terminating .END&E \\

\hline
Missing transistor model name&M20 2 3 0 0 (model undefined) &E\\
\hline
Floating node&in&E\\
\hline
Extra node in transistor definition&M20 2 3 2 0 \textbf{4} PMOS&M \\
\hline
Incorrect resistor value format&R1 in out \textbf{1K} (should be 1k)&M\\
\hline
Incorrect subcircuit definition&.SUBCKT a b (instance name undefined)&M \\
\hline
Missing capacitor value&C12 in out&M\\
\hline
Incorrect usage of transient analysis&.tr 100p (simulation duration undefined)&M\\
\hline

Missing voltage value &VB 5 0 & M\\
\hline
Special characters in node names&M\textbf{!2} 3 5 0 0 NMOS&D\\
\hline
Incorrect .PRINT statement&.PRINT TRAN \textbf{in} (should be \textbf{V(in)})&D \\
\hline
Incorrect current source definition&Ib 1 0 1M&D \\
\hline
Incorrect transistor name&M1 2 3 0 0 \textbf{NMOSC}& D\\
\hline
Missing .END in subcircuit&Subcircuit without .ENDS&D \\
\hline
Incorrect parameter definition&.PARAM \textbf{R1=}&D \\
\hline
Incorrect .OPTIONS definition&\textbf{.OPTION} POST (should be .OPTIONS)&D \\
\hline
Missing value in .DC&.DC vin 0.1 5 (missing increment value)&D \\

\hline

\end{tabular}
 \begin{tablenotes}
\item The syntactical bugs in the SPICE examples are indicated in \textbf{bold}. Different bug complexities are indicated by easy (E), medium (M), and difficult (D).
\end{tablenotes}
 \vspace{-0.2cm}
 \label{bugbench}

\end{table}

\begin{table}
\centering
\fontsize{7.8}{7.8}\selectfont
\caption{{Information provided to the LLM for syntactical bug and Trojan detection tasks.}}

    \label{compare}
    \begin{tabular}{|c| c|} 
    \hline
{Task} & Information provided\\
\hline

Bug Detection&SPICE netlist + syntax rules \\
\hline
\multirow{2}{*}{Trojan Detection}&SPICE netlist + corresponding simulation log files \\
&(current and voltage) + circuit specification \\
\hline
\end{tabular}
\vspace{-0.3cm}
\label{inputs}
\end{table}

 \begin{table*}
\centering
\caption{Performance of LLMs in identifying syntactical bugs across different circuit types.}
\fontsize{7.2}{7.2}\selectfont

 \begin{tabular}{|c|c|c|c|c|c|c|c|c|c|c|c|} 
     \cline{1-12}

Case& {Circuit Type} & \multicolumn{3}{c|}{Bugs inserted} &\multicolumn{3}{c|}{Bugs Detected (\%)} &\multirow{2}{*} {Bug }&\multicolumn{2}{c|}{FPR (\%)}&\multirow{2}{*}{Bug} \\ 
 \cline{3-8}
 \cline{10-11}
& &&&&&&&\multirow{2}{*}{Coverage (\%)}&Without&\textbf{With}&\multirow{2}{*}{resolved (\%)}\\
 
&&Easy&Medium&Difficult&Easy&Medium&Difficult&& Instructions&\textbf{Instructions}&\\
\hline
1&Common source amplifier&2&5&5&100&100&80&91.6&38.8&0&100\\
&(Resistive load)&&&&&&&&&&\\
\hline

2&Common source amplifier&3&4&6&100&75&66.7&76.9&28.5&7.1&90\\
&(Resistive and capacitive loads)&&&&&&&&&&\\
  
 \hline
3&NMOS Transistor&4&5&6&100&100&83.3&93.3&16.6&0&100\\

 \hline
4&Switched capacitor &4&4&7&100&100&71.4&86.6&23.5&5.8&100\\
 
 \hline
 5&Inverter&4&4&7&100&75&71.4&80&14.2&0&83.3\\
  
\hline
6&Differential amplifier &3&6&8&100&100&87.5&94.1&15.7&0&81.2\\

\hline

7&Current mirror  &4&6&8&100&83.33&75&83.3&21&6.25&86.6 \\
&(Both NMOS and PMOS)&&&&&&&&&& \\
\hline 
8&Current mirror &3&5&7&100&100&71.4&86.6&18.75&0&76.9 \\
&(Differential pair)&&&&&&&&&& \\
\hline
9&OPAMP &4&5&7&100&100&100&100&31.2&6.25&87.5\\

\hline
10&Bandgap filter &3&6&5&100&100&80&92.8&31.5&13.3& 92.3\\
\hline

\end{tabular}
\begin{tablenotes}
\item Using a refined set of SPICE syntax rules results in a significantly lower FPR compared to the scenario when a basic set of syntax rules is prompted to the LLM.
\end{tablenotes}
 \label{bugresults}

\end{table*}

\begin{table*}[t]
\centering
\caption{Performance of LLMs in detecting analog Trojans and locating Trojan-impacted nodes for several analog designs.}
\fontsize{7.5}{7.5}\selectfont

 \begin{tabular}{|c|c|c|c|c|c|c|c|c|c|c|c|c|c|c|c|c|c|} 
     \cline{1-18}

Case&Netlist&\multicolumn{4}{c|}{Trojan identified?}&\multicolumn{4}{c|}{Trojan Coverage (\%)}&\multicolumn{4}{c|}{Precision (\%)}&\multicolumn{4}{c|}{Recall (\%)}\\ 

\cline{1-18}
&&$FS$&$R$&\textbf{$R_{ FS}$}&$R^*_{ FS}$&$FS$&$R$&$R_{ FS}$&$R^*_{FS}$&$FS$&$R$&$R_{ FS}$&$R^*_{FS}$&$FS$&$R$&{$R_{FS}$}&$R^*_{FS}$ \\
\cline{3-18}
1&642\_troj\_1&\checkmark&\checkmark&\checkmark&\checkmark&57.14&14.28&\textbf{100}&\textbf{100}&67&67&\textbf{100}&\textbf{100}&67&67&\textbf{100}&\textbf{100}
\\
\hline
2&642\_troj\_6&\checkmark&\xmark&\checkmark&\checkmark&28.57&0&\textbf{100}&\textbf{100}&43&0&\textbf{100}&\textbf{100}&100&0&\textbf{100}&\textbf{100}
\\
\hline
3&642\_troj\_7&\xmark&\checkmark&\checkmark&\checkmark&0&14.28&100&100&0&25&67&67&0&33.3&67& 100
\\
\hline
4&642\_troj\_20&\checkmark&\checkmark&\checkmark&\checkmark&71.5&57.14&100&85.7&43&12.5&100&100&67&33.3&100&67
\\
\hline
5&642\_troj\_34&\checkmark&\checkmark&\checkmark&\checkmark&42.8&42.8&100&100&100&12.5&100&67&100&33.3&100&100\\
\hline
6&642\_troj\_36&\checkmark&\xmark&\checkmark&\checkmark&57.14&0&100&100&43&67&100&67&33.3&67&100&100\\
\hline
7&642\_troj\_38&\checkmark&\checkmark&\checkmark&\checkmark&57.14&14.28&100&100&50&67&67&75&67&67&67&100\\
\hline
8&642\_troj\_39&\xmark&\checkmark&\checkmark&\checkmark&0&28.57&85.7&85.7&0&33.3&100&100&0&33.3&100&100\\
\hline
9&755\_troj\_1&\checkmark&\xmark&\checkmark&\checkmark&14.28&0&100&100&50&0&100&67&33.3&0&100&100\\
\hline
10&755\_troj\_2&\checkmark&\xmark&\checkmark&\checkmark&14.28&0&100&85.7&43&0&75&75&67&0&100&100\\
\hline
11&755\_troj\_3&\checkmark&\checkmark&\checkmark&\checkmark&42.8&42.8&100&85.7&43&12.5&100&67&67&33.3&67&67\\
\hline
12&755\_troj\_5&\xmark&\checkmark&\checkmark&\checkmark&0&42.8&71.4&100&25&12.5&100&100&67&33.3&100&67\\
\hline
13&755\_troj\_6&\xmark&\checkmark&\checkmark&\checkmark&0&42.8&100&100&0&40&100&75&0&67&100&100\\
\hline
14&755\_troj\_9&\checkmark&\checkmark&\checkmark&\checkmark&57.14&57.14&100&100&50&33.3&100&100&67&33.3&100&67\\
\hline
15&755\_troj\_10&\checkmark&\checkmark&\checkmark&\checkmark&71.4&42.8&100&100&100&25&100&75&100&33.3&100&100\\
\hline
16&755\_troj\_11&\checkmark&\checkmark&\checkmark&\checkmark&57.14&42.8&85.7&100&71.4&0&100&100&33.3&0&100&100\\
\hline
17&755\_troj\_12&\checkmark&\checkmark&\checkmark&\checkmark&28.57&14.28&85.7&100&50&33.3&75&75&67&33.3&67&100\\
\hline
18&755\_troj\_16&\checkmark&\checkmark&\checkmark&\checkmark&57.14&14.28&71.4&85.7&67&0&75&100&67&0&100&100\\
\hline
19&755\_troj\_19&\checkmark&\xmark&\checkmark&\checkmark&28.57&0&\textbf{100}&\textbf{100}&25&0&\textbf{100}&\textbf{100}&33.3&0&\textbf{100}&\textbf{100}\\
\hline
20&755\_troj\_23&\checkmark&\checkmark&\checkmark&\checkmark&28.57&28.57&71.4&71.4&50&33.3&75&100&67&33.3&100&100\\
\hline
21&755\_troj\_24&\xmark&\xmark&\checkmark&\checkmark&0&0&100&100&0&0&75&75&0&0&100&100\\
\hline
22&738\_troj\_3&\checkmark&\checkmark&\checkmark&\checkmark&14.28&14.28&85.7&100&67&40&100&100&67&67&67&100\\
\hline
23&738\_troj\_4&\checkmark&\checkmark&\checkmark&\checkmark&42.8&28.57&100&85.7&67&33.3&75&100&67&33.3&67&67\\
\hline
24&738\_troj\_7&\checkmark&\xmark&\checkmark&\checkmark&57.14&0&100&100&25&0&100&67&33.3&0&100&67\\
\hline
25&738\_troj\_12&\checkmark&\xmark&\checkmark&\checkmark&57.14&0&71.4&100&67&20&100&100&67&33.3&100&100\\
\hline
26&738\_troj\_13&\checkmark&\xmark&\checkmark&\checkmark&57.14&0&85.7&71.4&40&0&75&75&67&0&100&100\\
\hline
27&738\_troj\_16&\checkmark&\checkmark&\checkmark&\checkmark&28.57&28.57&85.7&71.4&40&40&100&67&67&33.3&67&100 \\

\hline
28&738\_troj\_17&\checkmark&\xmark&\checkmark&\checkmark&71.4&0&100&85.7&0&0&75&100&0&0&100&100 \\
\hline
29&738\_troj\_23&\checkmark&\checkmark&\checkmark&\checkmark&28.57&28.57&\textbf{100}&\textbf{100}&25&33.3&\textbf{100}&\textbf{100}&33.3&33.3&\textbf{100}&\textbf{100} \\
\hline
30&738\_troj\_26&\checkmark&\checkmark&\checkmark&\checkmark&14.28&28.57&100&85.7&40&33.3&75&100&67&33.3&100&100 \\
\hline
\hline
\multicolumn{6}{|c|}{\textbf{Average}}&36.18&20.93&\textbf{93.32}&\textbf{93.32}&43.04&22.47&\textbf{90.3}&\textbf{86.46}&52.39&27.81&\textbf{92.3}&\textbf{93.4} \\
\hline

\end{tabular}
 \begin{tablenotes}
\item The proposed framework detects Trojan-impacted nodes as well as the Trojan circuit with 100\% accuracy and zero false positives for the \textbf{highlighted} cases. Across all evaluated scenarios, applying supervised-learning rules combined with few-shot learning yields higher average Trojan coverage, precision, and recall (marked in \textbf{bold}).
\end{tablenotes}
 \vspace{-0.2cm}
 \label{trojresults}

\end{table*}

\subsection{Performance Evaluation of GPT Model}
\vspace{-0.1cm}

\subsubsection{Syntactical Bug Detection}
\vspace{-0.1cm}

The bug detection results are shown in Table \ref{bugresults}. We observe that GPT-3.5 demonstrates higher accuracy and broader bug coverage across all evaluated SPICE netlists. Additionally, the proposed instruction-following approach results in fewer false positives for the evaluated bugs compared to the scenario when instructions were not explicitly provided in the prompt. 
\vspace{-0.1cm}
\subsubsection{Functional Bug (Trojan) Detection}
\vspace{-0.1cm}
To test the inherent capability of LLM to analyze structured prompts and to further improve its accuracy through well-crafted prompts based on supervised-learning rules and few-shot examples, we evaluate the following four test cases:

\begin{enumerate}[leftmargin=*,topsep=0pt]
    \item $FS$: We evaluate the LLM by providing only few-shot examples.
    \item $R$: We provide supervised-learning rules without any few-shot examples.
    \item $R_{FS}$: We provide supervised-learning rules followed by few-shot examples such that LLM can correlate the examples with the established rules, and use these examples to determine the Trojan-impacted nodes for a new test netlist based on these rules.
    \item $R^*_{ FS}$: We incorporate Rule \#2 instead of Rule \#1 (see Supervised Learning Rule 2), keeping the other rules unchanged, followed by few-shot examples.
\end{enumerate}

To increase the complexity for LLM evaluation, we scrambled the Trojan components and nodes within the netlist as well as changed the parameters, such as the width-to-length (W/L) ratios of transistors and the capacitor values. The Trojan-inserted netlists used for the experiments are labeled as `$netlist\_troj\_n$', where $netlist$ is the specific circuit chosen from the AMSNet repository \cite{amsnet} and $n$ indicates the netlist node where the Trojan payload is activated. Evaluation results are shown in Table \ref{trojresults}. We observe that incorporating the supervised-learning rules lead to a higher precision as well as overall accuracy in identifying the Trojan-impacted nodes. Additionally, the combination of few-shot examples with these rules enables the LLM to successfully identify the Trojan circuit in the benchmark netlists. The proposed method shows superior Trojan coverage and accuracy of Trojan-impacted nodes compared to scenarios where only few-shot examples are used without providing the supervised-learning rules.\par 
The maximum number of tokens fitting within the context window for the GPT-3.5-turbo model is 16385. Average LLM inferencing time is $9.2$ seconds across all cases of Trojan identification, thereby confirming the runtime efficiency of SPICED as it streamlines the process of analog design analysis and bug (both syntactical and functional) localization.

%% file: Conclusion.tex
\section{Conclusion}
\vspace{-0.1cm}

We have explored both syntactical and functional bug detection capabilities of the LLM. The proposed instruction-following approach significantly reduces the number of false positives while achieving high bug coverage. Additionally, by curating prompts with few-shot examples and CoT, LLM efficiently localizes the Trojan-impacted nodes for a range of Trojan-insertion scenarios while incurring zero area and power overheads. By incorporating the supervised learning rules in the prompt, we achieve an  average Trojan coverage of 93.32\% and an average true positive rate of 93.4\% in identifying Trojan-impacted nodes for the evaluated analog benchmark circuits. This opens up new directions for securing analog designs from threats arising anywhere between design and fabrication stages of the chip.